\documentclass{ws-ijmpa}

\begin{document}

\markboth{Keh-Fei Liu}{Pentaquark}

\catchline{}{}{}{}{}

\title{A Review of Pentaquark Calculations on the Lattice}

\author{Keh-Fei Liu}
\address{Dept. of Physics and Astronomy, University of Kentucky, Lexington, KY, 40506,
USA}

\author{Nilmani Mathur}
\address{Jefferson Lab., 12000 Jefferson Ave., Newport News, VA 23606, USA}  


\maketitle


\begin{abstract}

     We review lattice calculations of pentaquarks and discuss
issues pertaining to interpolation fields, distinguishing 
the signal of pentaquarks from those of the KN scattering states, chiral 
symmetry, and ghost state contaminations. 


\end{abstract}



\section{Introduction}

\hspace{0.5cm}   
 The recent interest in pentaquark baryons has been inspired by the
experimental discovery of $\Theta^+(1540)$ whose quark composition
is $uudd\bar{s}$. We will not address the experimental situation. It is 
summarized by T. Nakano in his plenary talk during this conference~\cite{nak05}.

 In the past few years, there has been quite an effort to
calculate these pentaquark states to see if they can be observed in 
lattice QCD calculations and if their masses and other properties agree 
with those observed experimentally. In particular, it would be interesting
to see if there is a way to understand why the observed width of $\Theta^+(1540)$
is much narrower than those of the ordinary baryons which are of the order of
several hundred MeV.

Somewhat similar to the experimental situation, among the dozen or
so lattice calculations, 5 calculations~\cite{cfk03,sas04,tuo05,ch05,at05a} 
claim to have positive signals for the pentaquark; while 7 calculations
~\cite{mla04,idi05,akt05,lhl05,cfk05,kj05,neg05} reported null results.
Prior to pentaquarks, lattice calculations dwell mostly on stable particles in  
strong decays, such as the nucleon, the pion, and kaon. Since the quark mass in
the present lattice calculations is still not light enough, the $\Delta$, $\rho$, 
and $\phi$ are also below the decay thresholds and can be calculated as the 
ground states. Come pentaquark, the interpolation field involves two $u$ quarks, 
two $d$ quarks, and one strange antiquark which inevitably will couple to the $KN$ 
scattering states in addition to the potential pentaquark state. This presents a 
challenge not confronted before and, presumably, posted some confusion in the first 
round of lattice calculations. We shall address some of the issues due to this complication, 
namely the questions of interpolation fields, the issue of $KN$ scattering states,
the importance of chiral symmetry, and the contamination of ghost states -
a quenched artifact and will attempt to draw some lessons from these
calculations.

\section{Interpolation Fields}   \label{IF}

\hspace{0.5cm}  
    There has been quite a lot of discussion in the literature on the choice 
of interpolation fields. The most naive one is the product of nucleon and kaon 
interpolation fields, i.e.
\begin{equation}
\chi_{1}^{I=0,1} =
\epsilon^{abc}\left(u^{Ta} C \gamma_5 d^b \right) 
\left[u^c \left( \bar{s}^e \gamma_5 d^e \right)
\mp \left\{u \leftrightarrow d\right\}\right],
\label{op1}
\end{equation}
where the $\mp$ combination is for $I=0$ and $I=1$ and has been denoted as such.. Another one
is similar, except with the color indices $e$ and $c$ positioned differently
\begin{equation}   \label{chi2}
\chi_{2}^{I=0,1} =
\epsilon^{abc}\left( u^{Ta} C \gamma_5 d^b \right)
\left[u^e \left( \bar{s}^e \gamma_5 d^c \right)
\mp \left\{u \leftrightarrow d\right\}\right],
\label{op2}
\end{equation}
The third one which has been used in the lattice pentaquark calculation is 
\begin{equation}
\chi_{3}^{\Gamma} = \epsilon^{gce}\epsilon^{gfh}\epsilon^{abc}
\left(u^{Ta} C \gamma_5 d^b \right) 
\left(u^{Tf} C\Gamma  d^h \right)\Gamma C^{-1} \bar{s}^{Te},
\end{equation}
with $\Gamma = \{S, A\} \equiv \{1, \gamma_{\mu}\gamma_5\}$~\cite{sas04}. It is motivated by 
the diquark-diquark-antiquark picture of Jaffe and Wilczek~\cite{jw04}.

The masses and spectral weights due to an interpolation field operator $O$ can
be extracted from the zero-momentum correlation function with a point source, 
which is a sum of exponentials for the spectrum of states with the quantum numbers of
the operator $O$
\begin{equation}
\langle \sum_{\vec{x}} O(\vec{x},t)\, O(\vec{x}_0,0)\rangle = \sum_i W_i\, e^{-m_i t}.
\end{equation} 
It is often asserted in the literature that $\chi_1$, being the $KN$ interpolation 
field, will couple to the $KN$ scattering states stronger than do $\chi_2$ and 
$\chi_3$. On the other hand, $\chi_3$ is expected to couple weakly to $KN$ states
and more strongly to the pentaquark state. However, it has been learned in the lattice
community over the years not to be overly sanguine about predicting the structure of 
the hadron based on the interpolation field. It is already shown in Ref.~[4] 
that $\chi_1, \chi_2$, and $\chi_3$ are linearly related through
the multiplication of $\gamma_5$ and a Fiertz transform between the u and $\bar{s}$
fields, i.e.
\begin{equation}
\gamma_5 \times (\chi_1^{I=0} - \chi_2^{I=0}) = \frac{1}{2}(\chi_3^S + \chi_3^A).
\end{equation}
Thus they are expected to couple to the same states ($1/2^-$ and $1/2^+$) with comparable 
spectral weights.

\begin{figure}
  \centerline{%
    \includegraphics[height=6.0cm,width=10.0cm]{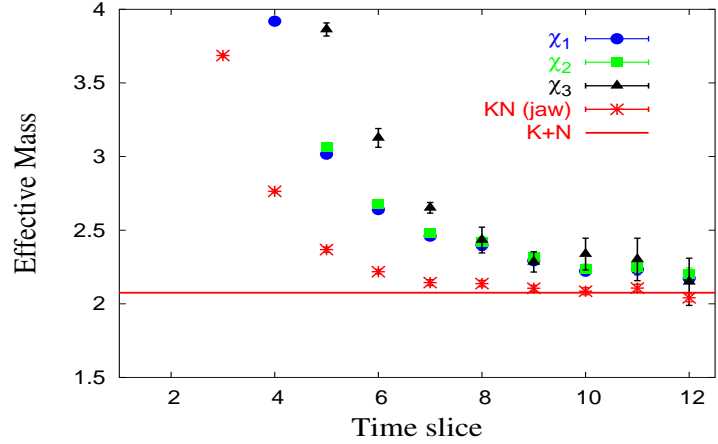}
  }
  \caption{The effective mass of the $1/2^-$ ground states from the $\chi_1, \chi_2, 
\chi_3$ and the $KN$ open jaw diagram as a function of time separation.  }
\end{figure}
\bigskip

\begin{figure}
  \centerline{%
    \includegraphics[height=4.0cm,width=6.0cm]{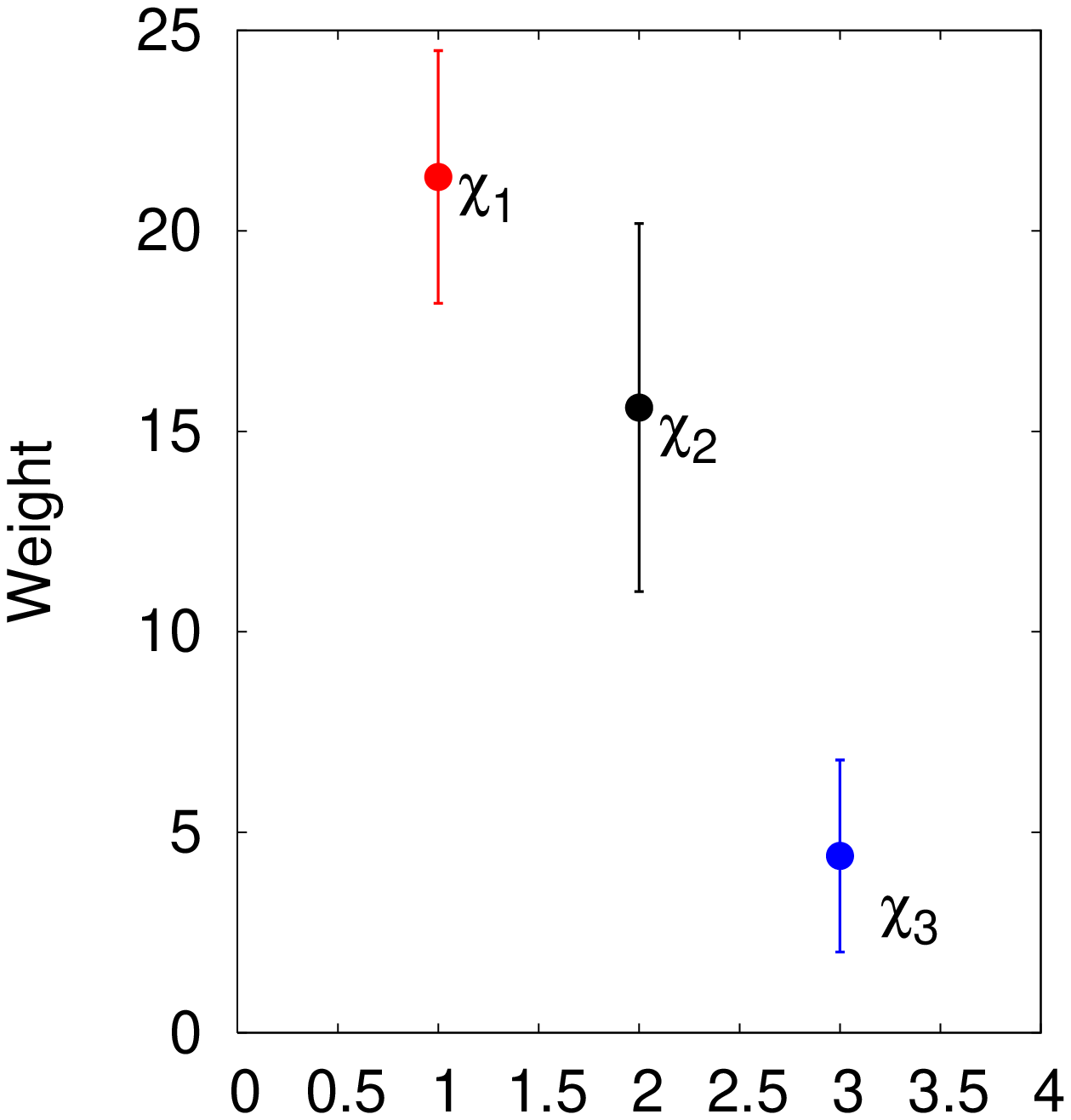}
    \includegraphics[height=4.0cm,width=6.0cm]{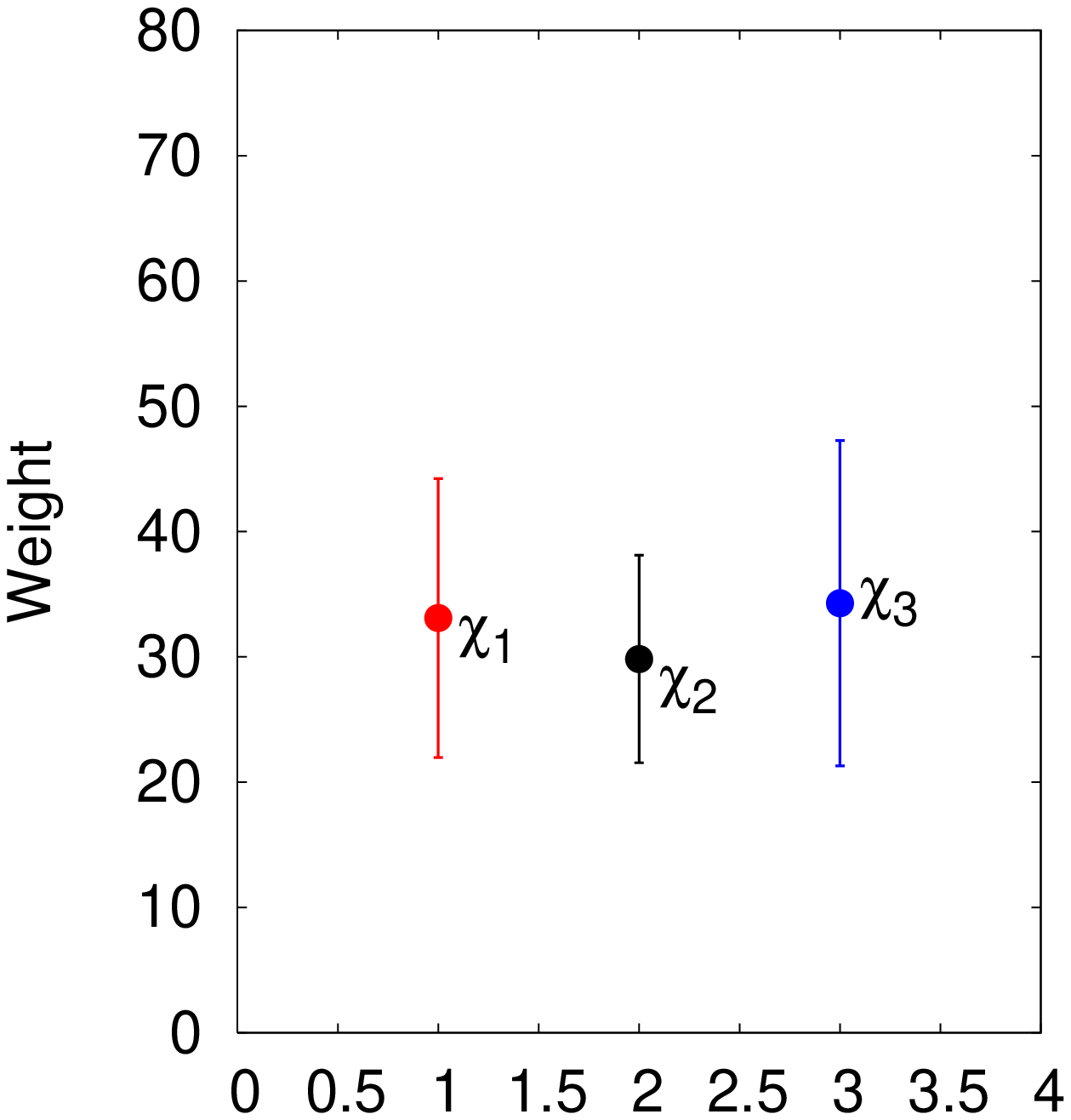}
  }
  \caption{The spectral weights (in arbitrary units) of the $1/2^-$ (left panel) and $1/2^+$ 
(right panel) ground states from the $\chi_1, \chi_2,$ and $\chi_3$ interpolation fields.}
\end{figure}

    As a further check, we show in Fig. 1 the effective mass plot of
the ground states as obtained with the three operators in the $1/2^-$ channel for
the $u/d$ quark mass corresponding to a pion mass at 633 MeV and the strange mass
corresponding to the physical $\phi$ mass on a $16^3 \times 28$ lattice with overlap 
fermions and a lattice spacing at $0.2$ fm~\cite{mla04}. We see that at short time 
separation, i.e. $t \leq 7 \, (1.4 {\rm fm})$, the three interpolation fields do not give 
the same mass. Only when $t \sim 12\, (2.4 {\rm fm})$ do they come down to the $KN$ threshold. 
Whereas the `open-jaw diagram', where the kaon and nucleon are separately projected to
the zero momentum states, has better overlap with the threshold $NK$ scattering state
and come down to the threshold much earlier. This shows that the three interpolation
fields project to the same ground state at large time separation. We also plot their
ground state spectral weights in the $1/2^-$ and the $1/2^+$ channels in Fig. 2. We see that
the spectral weight of $\chi_3$ is somewhat smaller than those of $\chi_1$ and $\chi$
in the $1/2^-$ channel ($KN$ in S-wave), but not orders of magnitude smaller; whereas,
all three give the same weights in the $1/2^+$ channel ($KN$ in P-wave). A single 
channel approach with enough time separation will be able to determine the ground
state reliably. But if the pentaquark state is close to the $KN$ scattering state,
as is considered in several calculations as a possbility in the $1/2^-$ channel, variational 
approach with multiple interpolation 
fields is a more effective approach. As a rule of thumb, the highest state will always 
be contaminated by still higher states and, thus, cannot be trusted. It is with the multi-operator 
variational calculations~\cite {cfk05,neg05} that one realizes that the earlier signal~\cite{cfk03}
of a low-lying $1/2^-$ state in addition to the $KN$ state at the threshold is spurious
and is due to the fact that there were only two operators in the variational calculation.

    It is also widely speculated that the diqaurk-diqaurk-antiquark interpolation field 
proposed by Jaffe and Wilczek~\cite{jw04} 
\begin{equation}
\chi_{JW}=\epsilon^{gfc}\epsilon^{def}\epsilon^{abc}(u^{Ta} C\gamma_5 d^b)\stackrel
{\leftrightarrow}{\cal{D}}_{\mu}(u^{Td} C\gamma_5 d^e)\gamma_{\mu}\gamma_5 
C \bar{s}^{Tg}
\end{equation} 
will have a better overlap with the pentaquark state to reflect the structure 
of the pentaquark. We note that this operator is very similar to the operator
$\chi_3^A$
\begin{equation}
\chi_3^A= \epsilon^{gfc}\epsilon^{def}\epsilon^{abc}(u^{Ta} C\gamma_5 d^b)
(u^{Td} C\gamma_{\mu}\gamma_5 d^e)\gamma_{\mu}\gamma_5 
C \bar{s}^{Tg}.
\end{equation}
The only difference is that there is a covariant derivative between the two diquarks in 
$\chi_{JW}$, while $\chi_3^A$ has a $\gamma_{\mu}$ between $u^T$ and $d$ in one of the $u-d$ 
diquark pair.

It turns out the pentaquark correlation functions built from these two interpolation fields
are linearly related. To see this, we consider the part of the correlator where the $\stackrel
{\leftrightarrow}{\cal{D}}_{\mu}$ and $\gamma_{\mu}$ operate on the d quark propagator to
the right which can be written in terms of the eigenstates of the Dirac
operator, i.e. $S(x,y) = \sum_{\alpha} \psi_{\alpha}(x)\psi_{\alpha}^{\dagger}(y)/
(i \lambda_{\alpha} + m)$. Now, the ${\cal{D}_{\mu}}$ and $\gamma_{\mu}$ operating 
on each eigenstate give
\begin{equation}
{\cal{D}}_{j} \psi_{\alpha} = \left( \begin{array}{c} {\cal{D}}_{j} U_{\alpha} \\
 {\cal{D}}_{j} L_{\alpha} \end{array} \right ); \,\,\,\, \gamma_{j} \psi_{\alpha} =
\left( \begin{array}{c} -i \sigma_j L_{\alpha} \\ i \sigma_j U_{\alpha} \end{array} 
\right ).
\end{equation}
The upper component of ${\cal{D}}_{j} \psi_{\alpha}$ and $\gamma_j \psi_{\alpha}$,\, i. e.
${\cal{D}}_{j} U_{\alpha}$ and $-i \sigma_j L_{\alpha}$ are related through the Dirac 
eigenvalue equation
\begin{equation} \label{Dic}
{\cal{D}}_{j} U_{\alpha} + i \epsilon_{jki}\sigma_i {\cal{D}}_{k} U_{\alpha} = (\lambda_{j} -
i {\cal{D}}_{4})\sigma_{j} L_{\alpha}.
\end{equation}
The lower components are similarly related. 
Thus, as far as the role of the interpolation fields is concerned and barring special exceptions, 
one would expect $\chi_{JW}$ to be equivalent to 
$\chi_3^A$ which is in term equivalent to $\chi_1$ and $\chi_2$. There is no \mbox{a priori} 
reason why one operator will preferentially project to a particular state, be it the pentaquark 
or the $KN$ scattering state. We should remark that the derivative operator $\chi_{JW}$ is 
motivated by the nonrelativistic picture and projects to the upper component of the Dirac
eigenstates in the $1/2^+$ channel; whereas, $\gamma_5 \chi_1, \gamma_5 \chi_2$, and $\chi_3$ 
utilize the lower components. For a large quark mass, $\sigma_j L_{\alpha}$ is approximately 
equal to ${\cal{D}}_j U_{\alpha}/m$ from Eq. (\ref{Dic}). In this case, the correlation function
due to a local operator is $O(m^2)$ smaller than that of the derivative operator. This
is the reason that the derivative operators are usually used to calculate the orbitally
excited states in heavy quarkonia. To illustrate this point, we plot in Fig. 3 the
spectral weight of the $S_{11}$ which is obtained from the lower component of
the nucleon correlation function with the $1 - \gamma_4$ projection to the negative parity
state~\cite{mcd05}. We see that the spectral weight does go down like $m^2$ in the range of 
pion mass from 250 MeV to 350 MeV. Thus, for heavy quarks, it is better to use the derivative
operator as the interpolation field and obtain the negative-parity $S_{11}$ state from the upper 
component of the $S_{11}$ correlator. For small quark masses, the lower component of
the nucleon correlator works for the $S_{11}$ state just as well. We believe the same is true 
with the pentaquark calculations. One shold be able to reach the same five-quark  states 
with the derivative operator, the $\chi_3$ operator, or equivalently the $\chi_1$ or $\chi_2$ 
operators with comparable spectral weights when the quark masses are close to their physical values.

\begin{figure}
  \centerline{%
    \includegraphics[height=7.0cm,width=10.0cm]{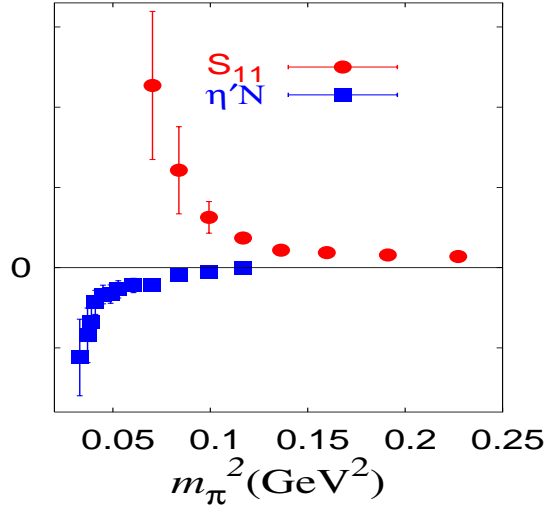}
  }
  \caption{The spectral weights (in arbitrary units) of $S_{11}$ and the $N\eta'$ ghost state
as a function of $m_{\pi}^2 ({\rm GeV}^2)$.  }
\end{figure}

\section{Distinguishing Pentaquarks from $KN$ Scattering States}

\hspace{0.5cm}  
Since the 5-quark interpolation field will, in general, project to both
the one-particle pentaquark states and the two-particle $KN$ scattering states on
the lattice, one needs to devise a way to distinguish them. To this end, it was advocated
to study the volume dependence of the spectral weights~\cite{mla04,mcd05}. If it is a one 
particle state, the spectral weight for the correlator constructed with point source
and zero momentum point sink has essentially no volume dependence. On the other hand,
if it is a two-particle scattering state with relatively weak interaction, it is inversely
proportional to the 3 volume from the normalization factor. The detailed 
study of such volume dependence with overlap fermions by comparing results from the
$16^3 \times 28$ and $12^3 \times 28$ lattices with pion mass $m_{\pi}$ as low as 
180 MeV revealed that the ground states of both the $1/2^-$ and $1/2^+$ channels are 
scattering states~\cite{mla04}. One needs to be careful here in view of the fact that it is more
subtle to obtain the spectral weight than the mass of a state. The former is more
sensitive to the fitting procedure and the length of time separation in the correlator. 
As it is shown recently that only when the time separation is large enough ($\sim 2.7$ fm) 
and with accurate data is the volume scaling of the spectral weight for the $I=2\,\, \pi\pi$ 
scattering state verified~\cite{at05b}. In the $KN$-pentaquark system that we are concerned 
with, the ground state mass is quite high compared to that of one hadron. Furthermore, the 
excited states of the system, which are the discrete $KN$ scattering states with $N$ and $K$ 
each with discrete lattice momentum in units of $2\pi/L$, are stacked up more compact than the 
radial excitation in the one hadron case when the lattice size is reasonably large (e. g. 
more than $\sim 2.4$ fm). Both of these factors require a large time separation or a more 
sophisticated fitting routine, such as fitting with Bayesian priors~\cite{lep02} or with a 
variational approach, in order to obtain the spectral weights reliably~\cite{mla04}. The study of 
$I=2\,\, \pi\pi$ in Ref.~[14] serves as a caveat for the authors' earlier work on 
pentaquark where they found the volume dependence of the spectral weight behaves like that of a 
one-particle state~\cite{at05a}. Another calculation~\cite{tuo05}, which observes little 
volume dependence of the spectral weight for the first excited state in the $1/2^-$ channel, 
employed variational method with two operators and used two exponentials to fit the spectral 
weights. As is widely known in fitting procedures, one cannot trust the results of the 
highest fitted state which is the second state in this case. As we pointed out in 
Sec.~\ref{IF}, the lesson learned from the variational calculations with 
multi-operators~\cite{cfk05} vs two operators~\cite{cfk03} should be taken to heart. 
As such, the results of this study~\cite{tuo05} should be taken with a grain of salt. 

Another way of distinguishing a scattering state from a pentaquark state is the clever 
idea of adopting the `hybrid boundary condition'~\cite{idi05} with the anti-periodic spatial 
boundary condition for the $u$ and $d$ quarks and periodic condition for the strange quark. 
This way, the energies of the $KN$ scattering states will be raised compared to those from the 
usual periodic boundary condition; while those of the $uudd\bar{s}$ pentaquarks will stay the 
same~\cite{idi05}. Using this technique, it is found that the state near the $KN$
threshold in the $1/2^-$ channel is a scattering state.

The third way to tell $KN$ scattering state and pentaquark state apart is to examine the
volume dependence of the spectrum~\cite{cfk05,tuo05,at05a,at05b}. The one-particle state 
is not expected to be sensitive to the lattice volume when it is large enough for the 
specific quark mass; whereas, a weakly interacting two-particle state with relative momentum 
$p$ will have a volume dependence since $p$ is in units of $2\pi/L$. This approach applies well 
to our study since the $KN$ interaction is weak and, as a result, it does not distort the 
discrete $KN$ spectrum much from the non-interacting one. But it requires high statistics 
in order to discern the volume dependence which is milder compared to the $1/V$ scaling
of the spectral weight.

\section{Chiral Symmetry and Ghost States}

\hspace{0.5cm}
   It is learned from studying the quenched chiral logs in $m_{\pi}, m_N$~\cite{cdd04}, 
and the meson cloud effect in Roper resonance and $S_{11}(1535)$~\cite{mcd05} that chiral
dynamics begins to play an important role in baryons when the pion mass is lower than
$\sim 300$ MeV which characterizes the chiral regime. It is logically possible that the 
pentaquark state exists only for very light quark masses. After all, the first prediction 
of the anti-decuplet pentaquark is based on the chiral Skyrme model~\cite{dpp97}. Since 
most of the lattice calculations are carried out outside this chiral regime, the conclusion
drawn from these calculations may not be relevant in this case. To have a definite answer
on the existence of the pentaquark will require calculations in the chiral regime with
dynamical fermions. When in the chiral regime, one needs to be concerned with
the presence of ghost states for the quenched approximation and the partially quenched
case when the sea quark and valence quark masses do not match. The relevant ghost state 
in our case is $KN\eta'$ with the would be $\eta'$ loop not fully developed to that of
a physical $\eta'$. Instead, it has a double $\pi$ pole which has the pion mass and
breaks unitarity by giving a negative contribution to the 5-quark correlator. The ghost
state is a practical problem in the $1/2^+$ channel in the chiral regime where the
$N,K$ and $\eta'$ can be in relative S-wave which becomes lower than the P-wave
$KN$ state at certain mass and gives rise to a negative correlator~\cite{mla04}. In this case,
the ghost state and the physical states have to be fitted together before drawing any
conclusion~\cite{mcd05}. After the  $KN\eta'$ ghost state is removed, it is found that
the lowest physical state is the $KN$ P-wave scattering state in the $1/2^+$ channel~\cite{mla04}.

   There is only one lattice calculation which claims to have detected a pentaquark in
the $1/2^+$ channel~\cite{ch05}. When extrapolated to the chiral limit, it
approaches the $KN$ threshold. This contradicts the findings in all the other calculations. 
We speculate that this could be due to the contamination of the $KN\eta'$ ghost state. Even though 
the quark masses are somewhat higher than the chiral regime, the fact that
the spatial volume is small in this study ($L = 1.8$ fm) makes it more susceptible to
the negative contribution from the the three-particle $KN\eta'$ ghost state which scales like
$1/V^2$. Since the would be $\eta'$ has the same mass as the pion, the ghost state is 
expected to reach the $KN$ threshold at the chiral limit. Another noticeable feature of 
the calculation is that the overlap fermion results are found to be higher than those of the Wilson 
fermion for $m_{\pi}$ heavier than $\sim 550$ MeV in the positive parity channel 
and $\sim 700$ MeV in the negative parity channel~\cite{ch05a}. In the $1/2^+$ case, the mass
difference can be as large as $\sim 300$ MeV. This needs an explanation. In contrast, ground 
state masses in the $1/2^-$ channel from 6 different calculations with both the overlap and 
Wilson fermions are plotted in Fig. 28 in Ref.~[8] and they pretty much lay on top 
of each other in the pion mass range from 420 MeV to 890 MeV.

\section{Conclusions}

\hspace{0.5cm}
   To summarize, we have learned several lessons from studying the 5-quark system which 
includes two-particle scattering states in the spectrum. First of all, the appearance of 
the interpolation field does not necessarily betray the structure of the hadron. 
To learn about the structure of a hadron, it is better to study the three- and four-point 
correlation functions.
Secondly, it is essential to discern the one- or two-particle nature of the observed
state before making a claim of having detected a pentaquark. 
In order to resolve states close by, it is necessary to perform variational 
calculation with more interpolation fields than the number of states of 
interest. Lastly, it is necessary to go down to small enough quark masses where
the chiral dynamics dominates and get rid of the quenched or partially 
quenched ghost state in the $1/2^+$ channel before one can be sure if the pentaquark
exists.

   After close examination of the existing lattice calculations, we conclude that there is no 
convincing evidence that any pentaquark state has been observed so far.

   Of course, `absence of evidence is not evidence of absence'~\cite{kj05}.
To finally settle the issue, one needs to carry out realistic dynamical fermion
calculations at physical quark masses, do a variational calculation
to study the spectrum and the volume dependence of the spectral weight, and remove  
the potential ghost states in the partially quenched case.


\section*{Acknowledgments}
This work is partially supported by USDOE grants DE-FG05-84ER40154 and DE-FG02-95ER40907.
The authors thank C. Alexandrou, T. Draper, Z. Fodor, M. Oka, and S. Sasaki for useful discussions.



\end{document}